\documentclass[aps,prb,reprint,groupedaddress,superscriptaddress,nofootinbib,floatfix]{revtex4-1}
\usepackage{mathrsfs}
\DeclareMathAlphabet{\mathpzc}{OT1}{pzc}{m}{it}
\usepackage{color}
\usepackage{graphicx}
\usepackage{dcolumn}
\usepackage{bm}
\usepackage{array}
\usepackage{multirow}
\usepackage{epstopdf}

\usepackage{amsmath}
\usepackage{tabulary}
\usepackage{rotating}
\usepackage{graphicx} 
\usepackage{lipsum} 
\usepackage{tabularx}

\usepackage{subfigure}
\usepackage{parskip}

\usepackage{tikz}
\usepackage{amsmath}
\usepackage{footmisc}

\usepackage{float}

\usetikzlibrary{shadings,patterns}
\usetikzlibrary{positioning, shapes.geometric}

\begin{document}
	\preprint{APS/123-QED}
	
	\title{High-performance linear-scaling electronic structure method via chromatic superposition states}
    \author{Zhikang Jiang}
    \altaffiliation{These authors contributed equally to this work.}
    \affiliation{State Key Laboratory of Quantum Functional Materials, School of Physical Science and Technology, ShanghaiTech University, Shanghai, 201210, China}
    \affiliation{Key Laboratory of Quantum Materials and Devices of Ministry of Education, School of Physics, Southeast University, Nanjing 211189, China}
    
    \author{Zhizhi Xiao}\altaffiliation{These authors contributed equally to this work.}
    \affiliation{State Key Laboratory of Quantum Functional Materials, School of Physical Science and Technology, ShanghaiTech University, Shanghai, 201210, China}
    \author{Mingfa Tang}
    \affiliation{State Key Laboratory of Quantum Functional Materials, School of Physical Science and Technology, ShanghaiTech University, Shanghai, 201210, China}
    \author{Weiyu Li}
    \affiliation{State Key Laboratory of Quantum Functional Materials, School of Physical Science and Technology, ShanghaiTech University, Shanghai, 201210, China}
    \author{Zhaoru Sun}
    \affiliation{State Key Laboratory of Quantum Functional Materials, School of Physical Science and Technology, ShanghaiTech University, Shanghai, 201210, China}
    \author{Ke Xia}
    \email{kexia@seu.edu.cn} 
    \affiliation{Key Laboratory of Quantum Materials and Devices of Ministry of Education, School of Physics, Southeast University, Nanjing 211189, China}
    \author{Youqi Ke}
    \email{keyq@shanghaitech.edu.cn} 
    \affiliation{State Key Laboratory of Quantum Functional Materials, School of Physical Science and Technology, ShanghaiTech University, Shanghai, 201210, China}
    \date{\today}

\begin{abstract}
We introduce a high-performance linear-scaling electronic structure method that employs chromatic superposition states (CSS) as a low-dimensional, high-fidelity representation, which can be orders of magnitude smaller than the full Hilbert space. Grounded in the system's finite correlation length, the CSS representation aggregates the uncorrelated orbitals into a single basis via a graph-coloring scheme, and is independent of the system size yet accurately preserves all sparse operators in solving the Kohn-Sham equations. The projection onto CSSs is efficiently computed by employing the block-Lanczos Krylov method which features high hardware efficiency and linear-scaling cost, enabling fast calculation of large-scale Kohn-Sham density matrix. 
We show that this method already outperforms previous linear-scaling density matrix purification method by more than one order of magnitude in computational speed at even small scale, while preserving high accuracy. The practical utility of the CSS method is demonstrated through molecular dynamics simulation of a 10000 $H_2O$, and self-consistent calculation of a 1-million $H_2O$ with modest resources. 
\end{abstract}
	\pacs{Valid PACS}
	\maketitle
The growing application of quantum mechanical calculations across materials science, chemistry and biology has intensified the demand for high-accuracy electronic structure (ES) modeling of large-scale systems.\cite{Ratcliff2017,Dawson2022} However, the steep scaling of conventional density functional theory (DFT), which typically increases as the cube of the number of electrons ($O(N_e^3)$)\cite{HohenbergPKohnWDFT1964,KohnWandShamLJDFT1965}, severely limits its applications to systems beyond a few hundred atoms, for example large biomolecules, nanostructured materials, interfaces and extended defects. To break this scaling barrier, linear-scaling (LS) DFT methods\cite{1999ReviewModernPhysics,Bowler2012} have emerged as a critical area of research, and are becoming indispensable for simulating large-scale systems.

LS-DFT methods exploit the inherent locality of electronic structure, as encapsulated in the “nearsightedness” principle.\cite{Kohn1996} These methods encompass diverse strategies, such as the divide-and-conquer technique\cite{YangPhysRevLett1991}, localized orbital formulations,\cite{Fattebert2008} and density matrix-based minimization and polynomial expansion algorithms,\cite{VanderbiltPhysRevB1993,VanderbiltPhysRevB1994,NiklassonPRB2002,SGoedeckerandMTeterPRB1995} etc. While significant progress has enabled simulations of systems comprising tens of thousands to millions of atoms,\cite{Bowler2010,FENG2024109135,IEEEYANG,GordonBell2024,Vetsch2025} practical implementations of LS methods remain hampered by
 large computational prefactors, system-size-dependent overheads, and hardware-level inefficiencies. For instance, density-matrix-based LS methods rely on iterative sparse-sparse matrix multiplications,\cite{NiklassonPRB2002,SGoedeckerandMTeterPRB1995} which is difficult to execute efficiently on modern CPU and GPU architectures.\cite{Suryanarayana2013,Artemov2021}
A central limitation of conventional LS approaches is that all sparse operators are fundamentally represented within the full Hilbert space, resulting in the irregular sparse data structures. 

In this work, we explore the intrinsic compressibility of the full Hilbert space to achieve a high-fidelity, dense representation of all sparse operators for solving Kohn-Sham equations. The foundation is the chromatic superposition state (CSS), which originated from the estimation of sparse Jacobi matrix using graph-coloring technique in Ref.\onlinecite{Coleman1983}. It was later used for estimating the diagonals of sparse operators,\cite{Bekas2007,Tang2012} showing promise for developing LS Kohn-Sham solver.\cite{Wang2018} Grounded in the finite correlation length of the sparse operators in Kohn-Sham system, the CSSs aggregates the spatially distant and uncorrelated atomic orbitals into a single basis, and can provide a low-dimensional representation independent of system size for ES calculations. Here, we combine the CSSs with a block-Lanczos projection method\cite{Golub1977,Tichy2025,Ozaki2006,KrylovSubspace,SLZhangKrylovSubspacediagonal2012} for efficiently computing the inverse square root of the overlap matrix $S^{-\frac{1}{2}}$, orthogonally transformed Hamiltonian $H'=S^{-\frac{1}{2}}HS^{-\frac{1}{2}}$ and density matrix $\hat{\rho}$ , thereby realizing a high-performace LS ES method.

To begin, we consider the non-interacting systems that features sparse density matrix $\hat{\rho}$ (the Fermi-Dirac operator for insulators and finite temperature metals),\cite{Mahanmanybodytheory} expressed by the Green's function (GF) as,
\begin{equation}\label{DMGF1}
\hat{\rho} = -\frac{1}{\pi} \text{Im}[\int f(z) G(z^{+})dz],
\end{equation}
where the retarded GF $G(z^{+}) = (z^{+}S - H)^{-1}(z^{+} =z + i\eta)$($\eta$ is positive infinitesimal), and $S$ and $H$ are the sparse overlap and Hamiltonian matrices defined in a set of complete localized bases, the Fermi–Dirac function  $f(z) = 1/{\left[\exp(\beta(z - \mu))+1\right]}$ ($\beta=1/k_BT$, $\mu$ is chemical potential). 
\begin{figure}[!htbp]
	 \centering
	\includegraphics[width=0.9\columnwidth]{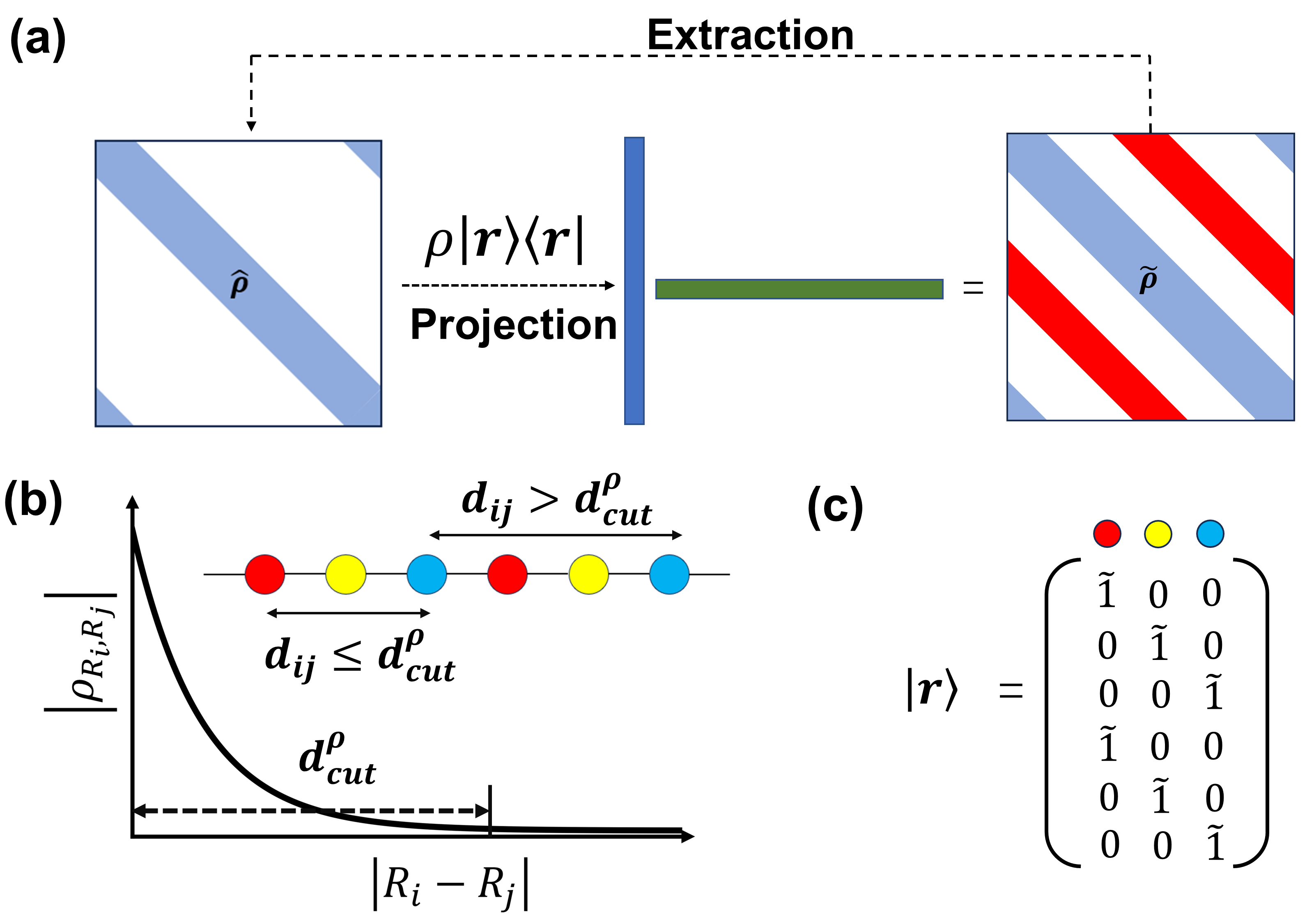}
	\caption{Schematic illustration of CSS method: (a) projection of sparse operators to CSS and extraction (blue for correct nonzero elements; red for contamination induced by CSSs); (b) graph-coloring method for the exponentially decaying density matrix, illustrated for a 1D atomic chain. (c) CSSs for the colored 1D chain in (b), with $\tilde{1}=\pm 1$.
 }
	\label{fig1}
\end{figure}

By revealing the sparsity of $\hat{\rho}$, we introduce a set of orthogonal CSSs ${\bf{r}}_c=\{|r_i\rangle, i=1,N^{\rho}_c\}$,\cite{Bekas2007,Tang2012} to provide a compressed representation of $\hat{\rho}$ with high fidelity, so that
\begin{equation}\label{eq-proj1}
\tilde{\rho} = \hat{\rho} \sum_{i=1}^{N^{\rho}_{c}}| r_i \rangle \langle r_i |=\hat{\rho}  |\bf{r}\rangle\langle\bf{r}|.
\end{equation}
To select the CSS to give correct non-zero elements of $\hat{\rho}$ as shown in Fig.\ref{fig1}(a), we here apply a greedy multicoloring algrithm\cite{Welsh1967} that assigns different colors (with total $N^{\rho}_{c}$ colors) for the spatial distance $d_{mn}$ of orbital indices m,n smaller than the cutoff distance $d^{\rho}_{cut}$,beyond which $\hat{\rho}_{mn}$ is negligible, as shown in Fig.\ref{fig1}(b). After coloring the system, for each color $i$ ($i\le N^{\rho}_c$), we construct a vector $| r_i \rangle$ by assigning $\pm 1$ to indeces colored $i$ while others are set to 0, as shown in Fig.\ref{fig1}(c). By setting appropriate $d^{\rho}_{cut}$, the accurate density matrix $\hat{\rho}$ can be extracted from the projection of Eq.\ref{eq-proj1}, namely $\tilde{\rho}$.  $d^{\rho}_{cut}$ reveals the finite correlation length in $\hat{\rho}$ and determines the number of CSSs $N^{\rho}_{c}$ which is thus independent of system size. The randomness in $\pm 1$ can be introduced to reduce the system error by stochastic cancellation. It is worth to mention that, different from CSS method exploiting the sparsity of density operators, the random-state methods ultilize the Monte-Carlo sampling and do the stochastic average.\cite{Tang_2024, PhysRevLett.111.106402, Zhou_2023}
We then rewrite $\tilde{\rho} = |\bf{X}\rangle\langle\bf{r}|$ with the $|X_i\rangle$ given by
\begin{equation}
|X_i\rangle = -\frac{1}{\pi} \text{Im} \int f(z) |R_i(z)\rangle dz,
\end{equation}
and $|R_i(z)\rangle=G(z)|r_i\rangle=(zS-H)^{-1}|r_i\rangle$ which is rewritten as, as important result of using GF in Eq.\ref{DMGF1},
\begin{equation}\label{Ri}
(z-H')|\tilde{R}_i(z)\rangle=|\tilde{r}_i\rangle, (i\le N_c^{\rho}),
\end{equation}
where $H'=S^{-\frac{1}{2}}HS^{-\frac{1}{2}}$, and $|R_i(z)\rangle=S^{-\frac{1}{2}}|\tilde{R}_i(z)\rangle$, $|\tilde{r}_i\rangle=S^{-\frac{1}{2}}|r_i\rangle$. Eq.\ref{Ri} forms a set of shifted linear systems, and shares a common Krylov subspace for different $z$. For multiple right handside of $|\tilde{r}_i\rangle$ ($i\le N_c^{\rho}$), we can construct the block-Krylov subspace of  dimension $v$ as  $\mathcal{K}^{v}(H',{\tilde{\bf{r}}})=span\{{\tilde{\bf{r}}},H'{\tilde{\bf{r}}}, H'^2{\tilde{\bf{r}}},..., H'^{v-1}{\tilde{\bf{r}}}\}$ via the block-Lanczos process to form an orthonormal space  $\mathcal{K}^{v}=\{|{\bf{\Phi}}^{\mathcal{K}}_m \rangle,m=1,...,v\}$.\cite{Golub1977}  Then, projection to $\mathcal{K}^{v}$ leads to the effective  Hamiltonian $\mathcal{H}^{\mathcal{K}}_{m,n}=\langle {\bf{\Phi}}_m^{\mathcal{K}}|H^{'}|{\bf{\Phi}}_n^{\mathcal{K}}\rangle$, with which we can write 
\begin{equation}
|\tilde{\bf{R}}(z)\rangle=|{\bf{\Phi}}^{\mathcal{K}}\rangle(z-\mathcal{H}^{\mathcal{K}})^{-1} \langle {\bf{\Phi}}^{\mathcal{K}} |\tilde{\bf{r}}\rangle,
\end{equation}
where $(z-\mathcal{H}^{\mathcal{K}})^{-1}$ is the GF in Krylov subspace. The density matrix in Eq.\ref{eq-proj1} can be thus rewritten as 
\begin{equation}
\tilde{\rho} = S^{-\frac{1}{2}} |\bf{\Phi}^{\mathcal{K}}\rangle \rho^{\mathcal{K}}\langle \bf{\Phi}^{\mathcal{K}}|\tilde{\bf{r}} \rangle \langle \bf{r}|,
\end{equation}
where
\begin{equation}\label{subspacedm}
\rho^{\mathcal{K}} = -\frac{1}{\pi} Im[\int{f(z)\left( z - \mathcal{H}^{\mathcal{K}}\right)^{-1}} dz].
\end{equation}
Due to the small size of $\mathcal{K}^{v}$, $\rho^{\mathcal{K}}$ can be efficiently computed by the exact diagonalization of $\mathcal{H}^{\mathcal{K}}$. The sparse density matrix $\hat{\rho}$ can thus be extracted from $\tilde{\rho}$ with high accuracy and efficiency. Compared to the approximate Chebyshev expansion of Fermi-Dirac functions,\cite{Bekas2007} present method treats it accurately in Eq.\ref{subspacedm}.

As an important step for computing $\hat{\rho}$, we obtain $S^{-\frac{1}{2}}$,which is a sparse matrix,\cite{Bowler2012} by projecting it onto $N_{c}^{S}$ colored probing states $|{\bf{o}}\rangle=\{|o_i\rangle,i=1,N_c^{S}\}$ sampled with appropriate $d_{cut}^{S}$, namely by computing $\tilde{S}^{-\frac{1}{2}}=S^{-\frac{1}{2}}|{\bf{o}}\rangle \langle\bf{o}|=|\bf{Y}\rangle\langle \bf{o}|$. With the relation 
\begin{equation}
S^{-\frac{1}{2}}=-\frac{1}{\pi}Im[\int z^{-\frac{1}{2}} G_S(z^+)dz],
\end{equation} 
where the GF $G_S(z^+)=(z^+ - S)^{-1}$, for solving $|\bf{Y}\rangle$, we can again construct a new Krylov subspace of dimension $v^s$ as  $\mathcal{K}_{S}^{v^s}(S,{\bf{o}})=span\{{\bf{o}},S{\bf{o}}, S^2{\bf{o}},...,S^{v^s-1}{\bf{o}} \}$ via the block-Lanczos process to form an orthonormal space  $\mathcal{K}_{S}^{v^s}=\{|{\bf{\Psi}}^{\mathcal{K}}_{S,m} \rangle,m=1,...,v^s\}$. Then, 
\begin{equation}
|{\bf Y}\rangle  =  |{\bf{\Psi}}_{S}^{\mathcal{K}}\rangle \mathcal{S}^{\mathcal{K},-\frac{1}{2}}\langle {\bf{\Psi}}_{S}^{\mathcal{K}} |\bf{o}\rangle,
\end{equation}

and 
\begin{equation}
 \mathcal{S}^{\mathcal{K},-\frac{1}{2}}= -\frac{1}{\pi} Im[\int{z^{-\frac{1}{2}}\left( z^{+} - \mathcal{S}^{\mathcal{K}}\right)^{-1}} dz] 
\end{equation}
where $\mathcal{S}^{\mathcal{K}}_{m,n}=\langle {\bf{\Psi}}_{S,m}^{\mathcal{K}}|S|{\bf{\Psi}}_{S,n}^{\mathcal{K}}\rangle$, and the integration is calculated by the exact diagonalization within Krylov subspace. Then, we obtain the sparse $H'$ by extracting from $S^{-\frac{1}{2}}H\tilde{S}^{-\frac{1}{2}}=S^{-\frac{1}{2}}H|\bf{Y}\rangle\langle\bf{o}|$. It is clear that, in all our computations including $S^{-\frac{1}{2}}$, $H'$ and $\hat{\rho}$, only sparse-dense matrix multiplications (SPMM) are involved, which is efficient in modern CPU and GPU architechtures.

To demonstrate the high accuracy and efficiency of the present block-Lanczos projected CSS method, we presently integrate it with density functional tight-binding method (DFTB)\cite{KaschnerPRB1995,FrauenheimIJOQC1996,FrauenheimPRB1998,MakinenComputationalMaterialsScience2009,SeifertPhysicalandEngineeringSciences2014} (within the open-source DFTB+ package\cite{Hourahine2020}) to self-consistently compute the ES, total energy and interatomic force, and benchmark against the exact diagonalization method (ED) and the second-order spectral projection (SP2) method (employing the NTpoly package\cite{Dawson2018}). DFTB employes the nonorthogonal and localized atomic basis set to provide good accuracy and transferability. In the present implementation, all the invovled sparse-dense matrix multiplications are performed parallely by using CRP-SPMM library.\cite{Huang2024} We simulate the large disordered $H_2O$ supercells with periodic boundary condition at $T=0$ with 6 atomic orbitals for each H2O. The behavior of water is governed by the delicate balance of complex interactions, e.g.covalent bond, hydrogen bond, weak dispersion interaction and the dipole interaction, and thus requires accurate calculation of density matrix, presenting a gold-standard system for testing linear-scaling methods. We use the DFTB-2 parameters of $H_2O$ system from the Refs.\onlinecite{Kullgren2017,Lourenco2020}. For the CSS-DFTB calculation of $H_2O$ molecules, $d_{cut}^{\rho}=14$ Bohr is used  to sample the CSSs $|\bf{r}\rangle$ with total $N_c^{\rho} \approx 310$ for $\hat{\rho}$ , and $d_{cut}^{S}=22$ Bohr to sample $|\bf{o}\rangle$ with $N_c^{S}\approx 560$ for both $S^{-\frac{1}{2}}$ and $H'$, with minor variations of $N_c^{\rho}$ and $N_c^{S}$ (within $\pm 20$) expected for different systems, to ensure the high accuracy in self-consistent calculation of density matrix.

\begin{figure}[!htbp]
	 \centering
	\includegraphics[width=0.9\columnwidth]{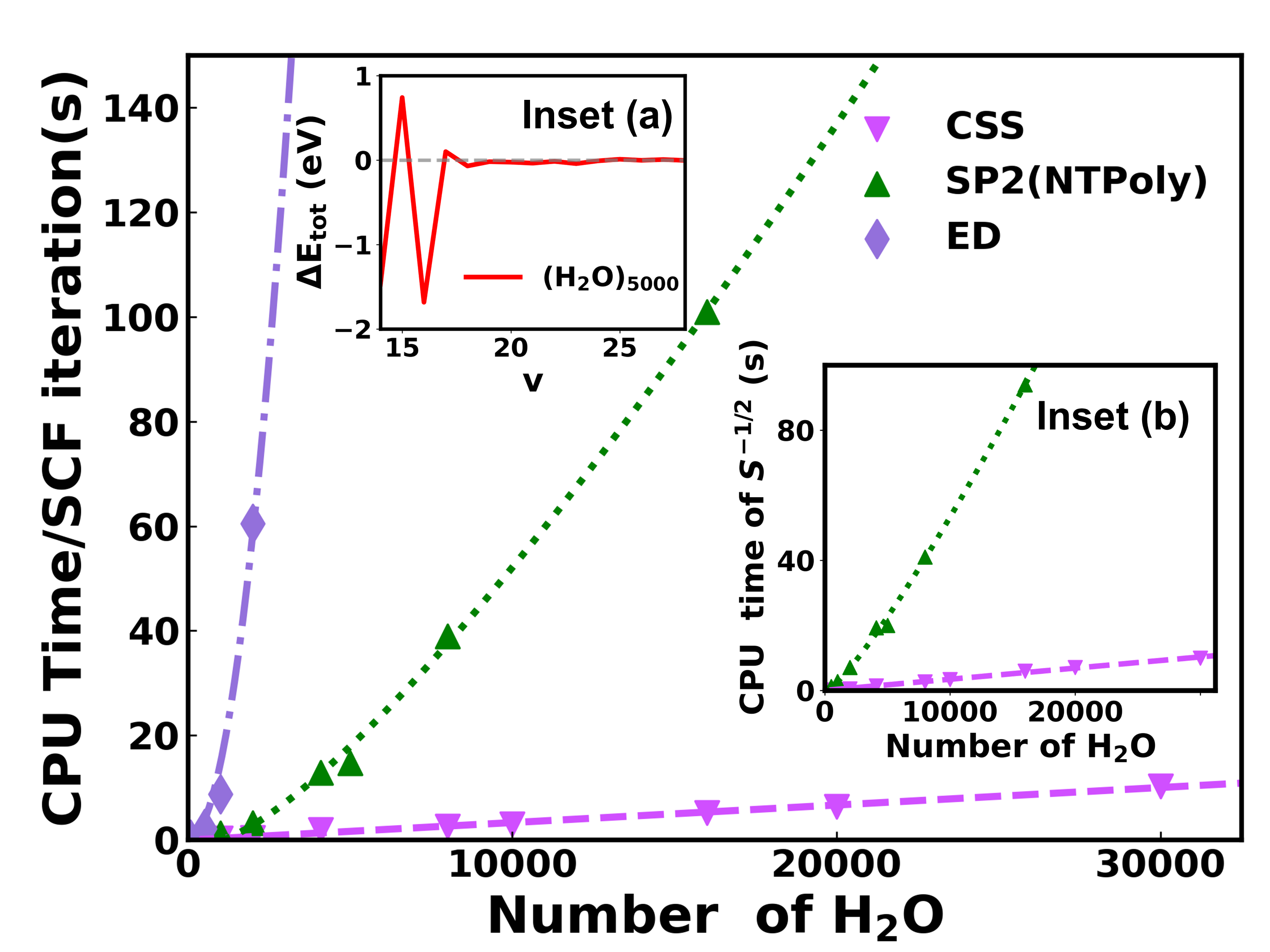}
	\caption{CPU time per SCF iteration versus number of $H_2O$ for different methods of CSS, SP2 (NTpoly) and ED. Inset(a): Deviation of $E_{orbit}$ versus the Krylov subspace $v$ for $(H_2O)_{5000}$. Inset(b): CPU time versus number of $H_2O$ for computing $S^{-1/2}$ for the SP2 and CSS methods. All the calculations are performed with 16 CPU cores of a single node.} 
	\label{fig2}
\end{figure}

In Fig.\ref{fig2} and the insets, we demonstrate the high computational efficiency of  CSS method by comparing with the SP2 and ED methods. All calculations with different methods implement a truncation threshold of $2*10^{-5}$ for maintaining matrix sparsity through selective element discarding, and are performed using 16-core parallelization on Intel Xeon Gold 6226 processors (2.70 GHz). As an important step before the self-consistent loops, both the CSS and SP2 methods compute $S^{-\frac{1}{2}}$ once for each atomic structure. Within the CSS method, $S^{-\frac{1}{2}}$ is obtained via block-Lanczos projection to CSSs, while, for the SP2, the NTpoly employs a Newton-Schultz iterative approach. Fig.\ref{fig2} Inset(b) compares the computational time of $S^{-\frac{1}{2}}$ for the CSS and SP2 method.It is clear that the present CSS method (with a small $v^s=10$ sufficient for converging the Krylov subspace) achieves a several-fold speedup compared to the Newton-Schultz algorithm in SP2. For example, for the $(H_2O)_{8000}$ and $(H_2O)_{16000}$, the CSS calculation takes the respective 3.4 and 6.8 seconds for $S^{-\frac{1}{2}}$, while the Newton-Schultz in SP2 takes the respective 39 and 94 seconds. 
Then, in Fig.\ref{fig2} Inset(a), we investigate the convergence behavior of total orbital energy $E_{orbit}=Trace(H\hat{\rho})$ versus the Krylov subspace dimension $v$ for a supercell $(H_2O)_{5000}$ at $T=0K$.
Remarkably, $E_{orbit}$ converges to within 18 meV accuracy at a modest $v=19$. This compact block-Krylov subspace guarantees the high computational efficiency essential for large-scale ES. Based on this convergence test, we employ $v=22$ throughout subsequent CSS calculations of $(H_2O)_{N}$. 

\begin{figure}[!htbp]
	 \centering
	\includegraphics[width=0.9\columnwidth]{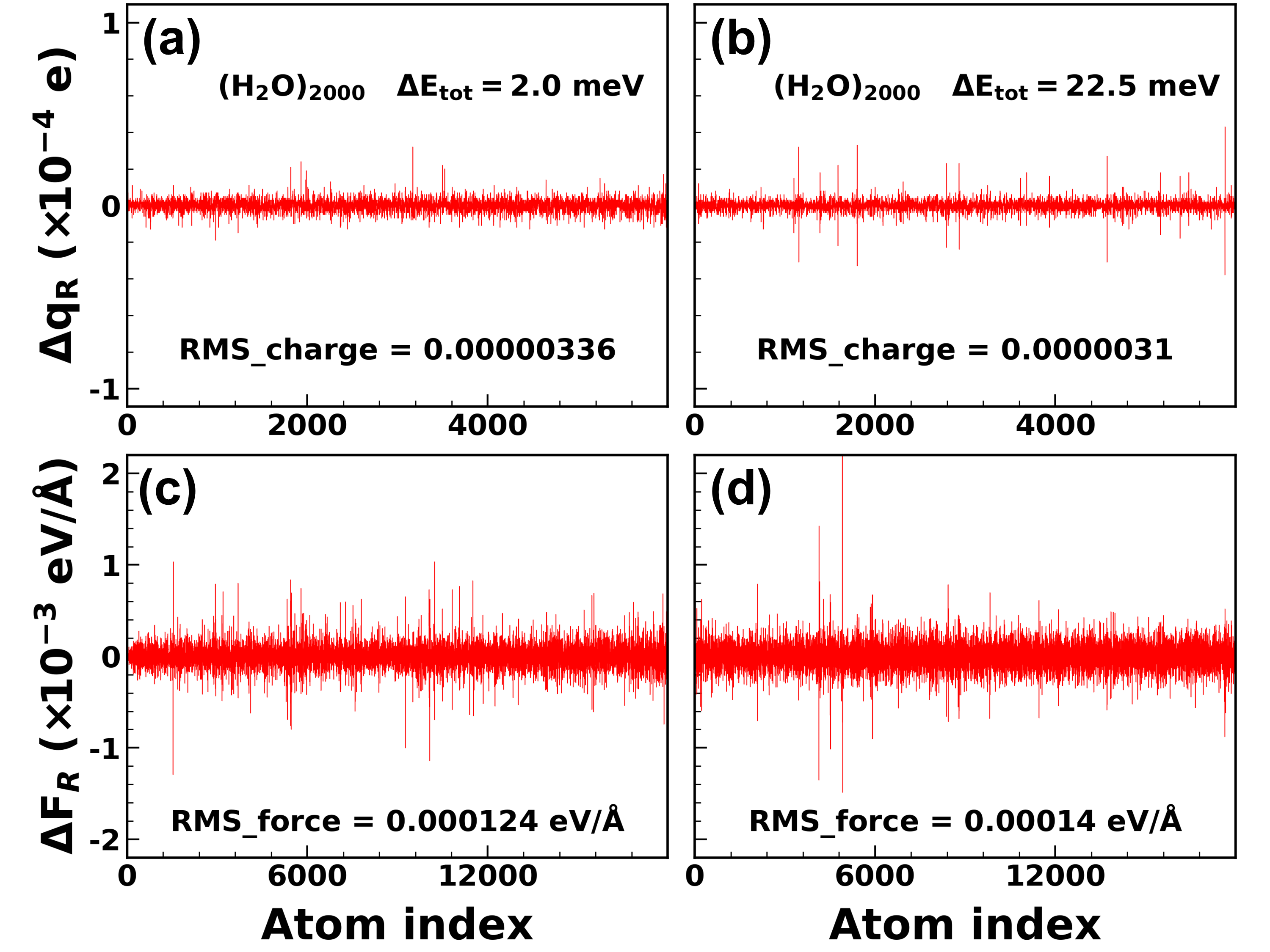}
	\caption{Deviation of Mulliken charge ($\Delta q_R$ in a,b) and force ($\Delta F_{R,x/y/z}$ in c,d) from the ED results for $(H_2O)_{2000}$: (a) and (c) are for the CSS method, (b) and (d) are for the SP2 method, total energy deviation, root-mean-square (RMS) errors of forces and charge deviation are also included.}
	\label{fig3}
\end{figure}

In Fig.\ref{fig2}, we compare the computational time (per self-consistent iteration) of diffferent methods for supercell $(H_2O)_{N}$ with sizes ranging from $N=64$ to 30,000. As shown, both the SP2 and CSS methods exhibit superior computational efficiency compared to the ED method which scales cubically ($N^3$).  It is observed that the SP2 algorithm’s scaling slightly deviates from linearity as system size $N$ increases, due to the fact that the iteration count in SP2 grows with system size as $ln(N)$.\cite{NiklassonPRB2002} In contrast, the CSS method maintains the linear scaling very well. It is important that the CSS method presents a significantly lower prefactor than SP2 in the computational time. For example, the CSS method takes 2.6 and 5.2 seconds per SCF iteration for the respective $(H_2O)_{8000}$ and $(H_2O)_{16000}$, more than one order of magnitude faster than the corresponding SP2 timings of 39 and 101 seconds.
The computational superiority of the block-Lanczos projected CSS method originates from the small number of sparse-dense matrix multiplication for constructing the Krylov subspace for key quantities $\hat{\rho}$, $S^{-\frac{1}{2}}$ and $H'$, while SP2 method relies on the sparse-sparse matrix multiplications which suffers from the hardware inefficiency and challenges on modern CPU and GPU architectures.

\begin{figure}[!htbp]
	 \centering
	\includegraphics[width=0.9\columnwidth]{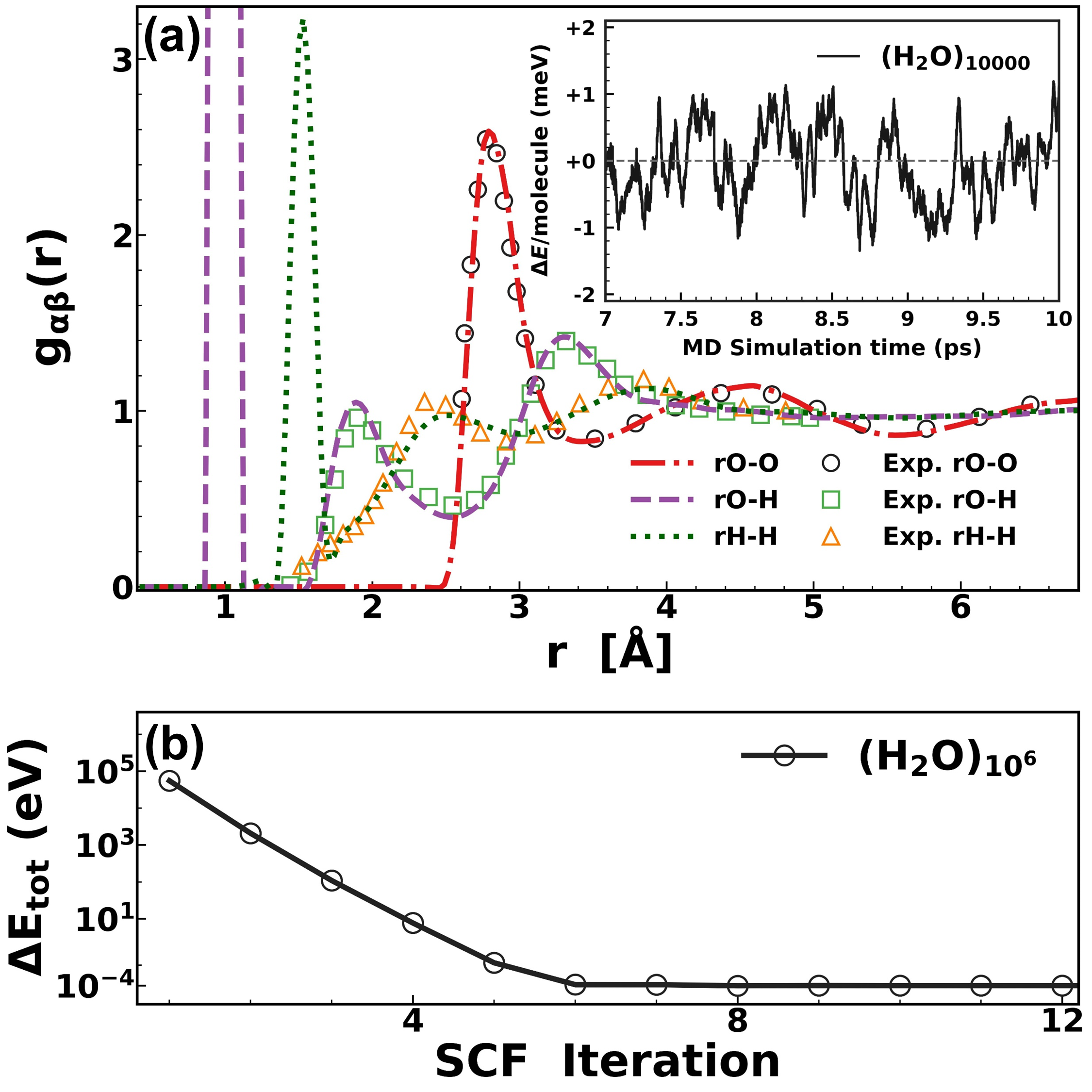}
	\caption{(a) Radial distribution functions $g_{\alpha\beta}(r)$ for O-O, O-H, and H-H pairs for MD simulations of bulk $(H_2O)_{10000}$ at 298K with the CSS method in DFTB, compared to the experimental measurements.\cite{Wikfeldt2009}(b) Total energy convergence versus SCF iterations for $10^6$ $H_2O$ system (shifted by the converged value of -111180290.941eV). The inset shows the MD total energy per $H_2O$ molecule (shifted by -111.106eV) versus MD simulation time. All the calculations are performed with 16 CPU cores of a single node }
	\label{fig4}
\end{figure}

In the next, we compare the accuracy of CSS method for the self-consistent ES calculations with the ED and SP2 results. The energy-weighted density matrix required for force calculations in DFTB, defined as $\hat{\rho}_{E}=\sum_{i=1}^{N}f(\epsilon_i)\epsilon_i|\psi_i\rangle\langle\psi_i|$ where $\psi_i$ and $\epsilon_i$ are the respective eigenstate and eigenvalue, is efficiently extracted by computing $\tilde{\rho}_E = S^{-1}H\tilde{\rho}$ in the CSS method. Furthermore, for efficient evaluation of the Coulomb potential and forces with the Mulliken charge, we employed the particle mesh Ewald (PME) method,\cite{Darden1993,Essmann1995} as implemented in helPME program,\cite{Simmonett2024} in place of the standard Ewald summation.
Fig.\ref{fig3} presents the deviations of Mulliken charge $\Delta q_R$ (in (a,b)) and force components $\Delta F_{R,x/y/z}$ (in (c,d)) from the ED results for both the CSS (a,c) and SP2 (b,d) calculations of $(H_2O)_{2000}$. 
As shown in Fig.\ref{fig3}, for both the deviations of Mulliken charge and force, the CSS results present the accuracy comparable to the SP2 calculations. Both SP2 and CSS methods present the maximum charge deviations $|\Delta q_R|_{max}\le 5*10^{-5}e$, and remarkably small root-mean-square errors (RMS) of $3*10^{-7} e$ for the $\Delta q_R$. Moreover, the force components computed by both methods exhibit the maximum deviations below 0.002 $eV/\AA$, with RMS errors of 0.000124 $eV/\AA$, demonstrating the reliability for large-scale molecular dynamics simulations. In addition, as shown in Fig.\ref{fig3}, the CSS method predicts the total energy deviation $\Delta E_{tot}=2 meV$ (from the ED calculation) for $(H_2O)_{2000}$, while the SP2 yields the $\Delta E_{tot}=22.5 meV$. For the high accuracy in the CSS calculations, besides the important choice of $d_{cut}^{\rho/S}$, the stochasticity in the random $\pm 1$ elements in the CSS construction (see Fig.\ref{fig1}) brings the self-cancellation effect for the errors in the charge, forces and total energy.

To demonstrate the practical utility and robustness of the CSS method for large-scale systems, we performed the molecular dynamics (MD) simulation of a water system. Specifically, a 10-ps NVT simulation was conducted for $(H_2O)_{10000}$ at room temperature (298 K) with the timestep of 0.5 fs. The radial distribution functions (RDFs) for O-O,H-O and H-H pairs were computed from the last 1 ps of the production trajectory, and compared with the experimental measurements, as shown in Fig.\ref{fig4}(a). All calculated RDFs agree well with experimental neutron diffraction data\cite{Wikfeldt2009} for the characteristic peak positions and coordination numbers of liquid water. For example, the CSS based DFTB reproduces the RDF peaks for O-O pair at $r_{O-O}=2.8,4.6\AA$, for O-H pair at $r_{H-O}=1.85,3.3\AA$ and for H-H pair at $r_{H-H}=2.5,3.9\AA$, illustrating the method's practical stability. To further demonstrate the scalability of the CSS method for large-scale systems, we conducted self-consistent ES calculations of a $10^6$ $H_2O$ supercell (using 16 cores of a single node). Fig.4(b) presents the total energy versus the self-consistent iterations. It is clear that the convergence can be rapidly reached after 6 iterations without the size-dependence, with the error magnitude reduced from $10^5$eV to 0.1 meV. The energy per water molecule (-111.180 eV) also shows remarkable agreement with small-scale benchmarks. Notably, each iteration requires only 644 seconds to construct the density matrix $\hat{\rho}$ (The deviation from ideal scaling arises from memory bandwidth limits of a single node). For comparison, a previous simulation of 1-million $H_2O$ using the divide-and-conquer method required 128000 CPU cores.\cite{Nishimura2018} Here for $10^6$ $H_2O$, the $N_c^{\rho}$ and $N_c^{S}$ are both more than 4 orders of magnitude smaller than the full Hilbert space of dimension $6\times10^6$. These results demonstrate the CSS method as an effective quantum mechanical approach for ultra-large scale materials simulations.

In summary, we have developed a highly efficient and accurate LS block-Lanczos projected CSS method for simulating large-scale electronic structure. This method projects the sparse operators/matrix function onto a small set of orthogonal CSSs generated by a graph-coloring technique with high fidelity. Combining with the block-Lanczos projection, the CSS method enables the efficient computation of $S^{-\frac{1}{2}}$, the transformed Hamiltonian $H'=S^{-\frac{1}{2}}HS^{-\frac{1}{2}}$ and the Fermi-Dirac operator. By integrating with DFTB, we demonstrate that the CSS method achieves remarkable computational efficiency and accuracy for large scale ES calculation, compared to the ED and SP2 calculations. The practical utility and robustness of the method are further illustrated by MD simulation of 10000 $H_2O$ and self-consistent calculation of $10^6$ $H_2O$ with modest computational resources. The CSS method offers an effective and scalable approach for quantum mechanical simulations of large scale materials.

Y.Ke acknowledges financial support from NSFC (grant No.12227901), and K.Xia thanks the support from the National Key Research and Development Program of China(GrantNos.2023YFA1406600),Z.Sun Thanks the support from Shanghai Rising-star Program (23QA1406800). The authors thank the HPC platform of ShanghaiTech University for providing the computational facility.


\begin{thebibliography}{10}


\bibitem{Ratcliff2017} 
L. E. Ratcliff \textit{et al.}, WIREs Comput. Mol. Sci. \textbf{7}, e1290 (2017).

\bibitem{Dawson2022} 
W. Dawson, A. Degomme, M. Stella, T. Nakajima, L. E. Ratcliff, and L. Genovese, WIREs Computational Molecular Science \textbf{12}, e1574 (2022).


\bibitem{HohenbergPKohnWDFT1964}
P. Hohenberg and W. Kohn, \emph{Phys. Rev}. 136, B864(1964).

\bibitem{KohnWandShamLJDFT1965}
W. Kohn, and L. J. Sham, \emph{Phys. Rev}. 140, A1133(1965).

\bibitem{1999ReviewModernPhysics}
S. Goedecker, Rev. Mod. Phys. 71, 1085 (1999).

\bibitem{Bowler2012} 
D. R. Bowler and T. Miyazaki, \emph{Reports on Progress in Physics}. 75, 036503 (2012).

\bibitem{Kohn1996} 
W.\ Kohn, \emph{Phys.\ Rev.\ Lett}.\ \textbf{76}, 3168 (1996).

\bibitem{YangPhysRevLett1991}
W. Yang,  \emph{Phys. Rev. Lett}. 66, 1438 (1991).

\bibitem{Fattebert2008}
J-L Fattebert, \emph{J.\ Phys.: Condens.\ Matter} \textbf{20}, 294210 (2008).

\bibitem{VanderbiltPhysRevB1993}
X.P. Li, R. W. Nunes, and D. Vanderbilt, \emph{Phys. Rev. B}. 47, 10891 (1993). 

\bibitem{VanderbiltPhysRevB1994}
R. W. Nunes and D. Vanderbilt, \emph{Phys. Rev. B}. 50, 17\,611 (1994). 

\bibitem{NiklassonPRB2002}
A. M. N. Niklasson. \emph{Phys. Rev. B}. 66, 155115(2002).

\bibitem{SGoedeckerandMTeterPRB1995}
S. Goedecker, M. Teter, \emph{Phys. Rev. B}. 51, 9455(1995)

\bibitem{FENG2024109135} 
J. Feng, L. Wan, J. Li, S. Jiao, X. Cui, W. Hu, and J. Yang, \emph{Computer Physics Communications} \textbf{299}, 109135 (2024).

\bibitem{IEEEYANG} 
Q. Jiang, Z. Cao, J. Chen, X. Qin, W. Hu, H. An, and J. Yang, \emph{IEEE Transactions on Parallel and Distributed Systems} \textbf{36}(7), 1495-1508 (2025).

\bibitem{Bowler2010} 
D. R. Bowler and T. Miyazaki, \emph{J. Phys.: Condens. Matter}\ \textbf{22},\ 074207\ (2010).

\bibitem{GordonBell2024} 
R. Stocks, J. L. G. Vallejo, F. C. Y. Yu, C. Snowdon, E. Palethorpe, J. Kurzak, D. Bykov, and G. M. J. Barca, in \emph{Proc. SC'24: International Conference for High Performance Computing, Networking, Storage and Analysis} (IEEE, 2024), pp. 1-12.

\bibitem{Vetsch2025} 
N. Vetsch, A. Maeder, V. Maillou, A. Winka, J. Cao, G. Kwasniewski, L. Deuschle, T. Hoefler, A. N. Ziogas, and M. Luisier, \emph{Proceedings of the International Conference for High Performance Computing, Networking, Storage and Analysis (SC'25)} (ACM, 2025), pp. 1-13.

\bibitem{Suryanarayana2013}
P.\ Suryanarayana, \emph{Chem.\ Phys.\ Lett}.\ \textbf{555}, 291 (2013).

\bibitem{Artemov2021}
A.\ G.\ Artemov and E.\ H.\ Rubensson, \emph{J.\ Comput.\ Phys}.\ \textbf{438}, 110354 (2021).

\bibitem{Coleman1983} 
T. F. Coleman and J. J. Moré, \emph{SIAM J. Numer. Anal.} 20, 187–209 (1983).

\bibitem{Bekas2007}
C.\ Bekas,\ E.\ Kokiopoulou,\ Y.\ Saad,\ \emph{Appl.\ Numer.\ Math}\ {\bf57},\ 1214\ (2007).

\bibitem{Tang2012}
J.M.\ Tang,\ Y.\ Saad,\ \emph{Numer.\ Linear Algebra Appl}\ {\bf19},\ 485\ (2012).

\bibitem{Wang2018} 
Z. Wang, G.-W. Chern, C. D. Batista, and K. Barros, \emph{J. Chem. Phys.} \textbf{148}, 094107 (2018).


\bibitem{Golub1977} 
G. H. Golub and R. Underwood, \emph{Mathematical Software, III (Proc. Sympos., Math. Res. Center, Univ. Wisconsin, Madison, Wis., 1977)}. pp. 361–377 (1977).

\bibitem{Tichy2025} 
P. Tichý, G. Meurant, and D. Šimonová, \emph{Numerical Algorithms} (2025).

\bibitem{Ozaki2006} 
Taisuke Ozaki,\ \emph{Phys.\ Rev.\ B}\ {\bf74},\ 245101\ (2006).

\bibitem{KrylovSubspace}
T. Sogabe, \emph{Krylov Subspace Methods for Linear Systems}.(Springer Nature Singapore Pte Ltd, Singapore, 2022).

\bibitem{SLZhangKrylovSubspacediagonal2012}
T. Hoshi and S. Yamamoto and T. Fujiwara and T. Sogabe, and S-L. Zhang, \emph{Journal of Physics: Condensed Matter}. 24, 165502 (2012).

\bibitem{Mahanmanybodytheory}
G. D. Mahan, \emph{Many-Particle Physics} (Springer Science Business Media, New York, 2013).

\bibitem{Welsh1967} 
D. J. A. Welsh and M. B. Powell, \emph{The Computer Journal}. 10, 85-86 (1967).

\bibitem{Tang_2024} 
M. Tang, C. Liu, A. Zhang, Q. Zhang, J. Zhai, S. Yuan, and Y. Ke, \emph{Chin. Phys. Lett.} \textbf{41}, 053102 (2024).

\bibitem{PhysRevLett.111.106402} 
R. Baer, D. Neuhauser, and E. Rabani, \emph{Phys. Rev. Lett.} \textbf{111}, 106402 (2013).

\bibitem{Zhou_2023} 
W. Zhou and S. Yuan, \emph{Chin. Phys. Lett.} \textbf{40}, 027101 (2023).


\bibitem{KaschnerPRB1995} 
D. Porezag, T. Frauenheim, T. Kohler, G. Seifert,and R. Kaschner,\emph{Phys. Rev. B}. 51, 12 947 (1995).

\bibitem{FrauenheimIJOQC1996}
G. Seifert, D. Porezag, and T. Frauenheim, \emph{International Journal of Quantum Chemistry}. 58, 185 (1996).

\bibitem{FrauenheimPRB1998}
M. Elstner, D. Porezag, G. Jungnickel, J. Elsner, M. Haugk, T. Frauenheim, S. Suhai and G. Seifert, \emph{Phys. Rev. B}. 58, 7260 (1998).

\bibitem{MakinenComputationalMaterialsScience2009}
P. Koskinen and V. Makinen, \emph{Computational Materials Science}. 47, 237(2009).

\bibitem{SeifertPhysicalandEngineeringSciences2014}
M. Elstner and G. Seifert, \emph{Philosophical Transactions of the Royal Society A: Mathematical, Physical and Engineering Sciences}. 372, 20120483 (2014).

\bibitem{Hourahine2020} 
B. Hourahine, et.al., \emph{J.Chem.Phys} {\bf152}, 124101(2020).


\bibitem{Dawson2018}
William Dawson,\ Takahito Nakajima,
\emph{\ Comp Phys Commun}\ {\bf225},\ 154--165,(2018).


\bibitem{Huang2024} 
H. Huang and E. Chow, \emph{IEEE Transactions on Parallel and Distributed Systems} \textbf{35}, 1977--1988 (2024)

\bibitem{Kullgren2017}
Jolla Kullgren,\ Matthew J Wolf, Kersti Hermansson, Christof Köhler, Bálint Aradi, Thomas Frauenheim, Peter Broqvist,
\emph{\ J Phys Chem C}\ {\bf121},\ 4593--4607,(2017).

\bibitem{Lourenco2020}
M. P. Louren{\c{c}}o, E. C. dos Santos, L. G. M. Pettersson, and H. A. Duarte, 
\emph{Journal of Chemical Theory and Computation} \textbf{16}, 1768--1778 (2020).


\bibitem{Darden1993} 
T. Darden, D. York, and L. Pedersen, \emph{The Journal of Chemical Physics} \textbf{98}, 10089--10092 (1993).

\bibitem{Essmann1995} 
U. Essmann, L. Perera, M. L. Berkowitz, T. Darden, H. Lee, and L. G. Pedersen, \emph{The Journal of Chemical Physics} \textbf{103}, 8577--8593 (1995).

\bibitem{Simmonett2024}
A. C. Simmonett et al., \emph{helPME}. GitHub (2024). \url{https://github.com/andysim/helpme}.  


\bibitem{Wikfeldt2009} 
K. T. Wikfeldt, M. Leetmaa, M. P. Ljungberg, A. Nilsson, and L. G. M. Pettersson, \emph{The Journal of Physical Chemistry B} \textbf{113}, 6246--6255 (2009).

\bibitem{Nishimura2018} 
Y. Nishimura and H. Nakai, \emph{Journal of Computational Chemistry}. 39, 105–116 (2018).


\end{thebibliography}

\end{document}